\documentclass[aps,onecolumn,12pt,showpacs,preprintnumbers]{revtex4}

\usepackage{amsmath}
\usepackage{graphicx}
\usepackage{epstopdf}

\renewcommand{\Im}{\mathop{\rm Im}}

\begin{document}

\title{Neutrino Oscillations and Decoherence}

\author{Luca Visinelli}
\email[Electronic address: ]{lucavi@physics.utah.edu}
\author{Paolo Gondolo}
\email[Electronic address: ]{paolo@physics.utah.edu}
\affiliation{The University of Utah,
Physics Dept. 115 S 1400 E, Salt Lake City, UT USA 84112}
\date{\today}

\begin{abstract}
We present an expression for the transition probability between
Dirac or Majorana neutrino flavors obtained from first principles within quantum field theory. Our derivation is based on a standard quantum
mechanical setup and includes the
specific mechanism of neutrino production only in as much as it specifies the initial state. Our
expression for the transition probability reproduces the usual
formula in the plane wave limit and shows the correct
non-relativistic and ultra-relativistic behaviors. It also allows a
simple understanding of the decoherence of the oscillations and of
the question of the arrival times of the different neutrino mass eigenstates. We show numerical examples for the case of two neutrino generations.
\end{abstract}



\maketitle

\section{Introduction}

Neutrinos are hoped to enlighten the search for new physics at both
the cosmological and microscopic level. The discovery of neutrino
oscillations at the SuperKamiokande \cite{SKK} and SNO \cite{SNO1, SNO2}
detectors prompted a detailed examination of the theory of neutrino oscillations (some early references are Pontecorvo~\cite{pontecorvo}, Kayser~\cite{kayser}, Bilenky~\cite{BP}; see Giunti and Kim~\cite{giunti1} for further references).

Some papers were dedicated to obtaining the formula for the
oscillation probability from first principles. There are some
aspects in these derivations that we do not find very precise and
that we want to address in this paper.

Beuthe \cite{beuthe} and Giunti \cite{giunti} stress the fact that
the pattern of neutrino oscillations depends on the mechanism of
production and detection, in other words, the oscillation amplitude
is different if neutrinos are produced in pion decay or from a
charged current interaction. We believe that, following standard
rules of quantum mechanics, the pattern of oscillations should only
depend on the initial state of the system. The particular production
mechanism needs to be specified only to determine the initial state.
Analogously, for detection, quantum mechanics can provide the
probability of detecting a neutrino in any specific final state. In
other words, the theoretical description of the phenomenon of
oscillations should involve the initial state, its evolution and a
final state representing the detection process. This is because of
the quantum mechanical properties of the system, which state that
different ways of producing the same initial wave packet should
bring to the same result.

In their articles, Beuthe \cite{beuthe} and Giunti \cite{giunti} average their final formulas for the oscillation probability over time, justifying this because the arrival time is not detectable. We will not perform this average because, from a fully theoretical treatment, oscillations occur both in space and time on the same footage. We will show how this feature of neutrino oscillations is related to the covariance of the oscillation formula and to the problem of defining an equal energy or equal momentum condition for the various mass eigenstates.

Snellman and Pallin \cite{pallin} derived the expression for the
oscillation amplitude by propagating a second-quantized neutrino
wave packet in vacuum and in matter, intentionally neglecting any
theory concerning production and detection of neutrinos. However in their paper there is a misuse of flavor and mass eigenstates. In fact, for the construction of the wave packets, the authors in \cite{pallin} use the neutrino flavor representation and introduce
concepts such as the energy $E_{\alpha}$ of a neutrino with flavor
$\alpha$. Since neutrino flavor states are not eigenstates of the
free Hamiltonian, the flavor-eigenstate energy $E_{\alpha}$ is not
well defined. Moreover, the authors in \cite{pallin} use the formula
$E_{\alpha} = \sqrt{p^{2}+m^{2}_{\alpha}}$, where they state that
$m_{\alpha}$ is the mass of the flavor $\alpha$. This is misleading
because a flavor eigenstate, being a linear combination of different
mass eigenstates, does not have a well-defined mass.

In this paper, we recover the formula for neutrino
oscillations in flat space-time using the propagation of second
quantized Gaussian wave packets. We choose wave packets instead of
monochromatic plane waves because we want to represent neutrino states that are ``localized" near the production and later near the detection point, but plane waves are ``everywhere" at the same time. In addition, wave packets are normalizable while plane waves are not. Another reason is that by using them we can
study the coherence between eigenfunctions with different mass (and
thus different group velocities). We also remedy the misuse of flavor states and flavor
properties in \cite{pallin} by deriving the neutrino oscillation formula using mass
eigenstates only. After all, it is the mass eigenstates that appear
in the perturbative expansions of the scattering processes, for
example in the Green functions and the Feynman rules.

Finally, we show numerical results for the case of two neutrino
generations, for realistic mass differences and for the pedagogical
case of a large mass difference. The latter case illustrates how
decoherence suppresses the oscillation phenomenon.

\section{Formalism}

Before recovering the transition probability formula we first establish our notation by reviewing
the description of the Dirac and Majorana free fields in Quantum
Field Theory in the Heisenberg picture. We also review the wave
packet formalism to be used later on.

A Dirac field $\hat{\psi}_{i}(x)$ describing a particle of mass
$m_{i}$ is \cite{srednicki}
\begin{equation} \label{dirac_field}
\hat{\psi}_{i}(x) = \sum_{\zeta}\int d\tilde{p}_{i}[\hat{b}_{i}({\bf
p},\zeta)u({\bf p},\zeta)e^{-ip \cdot x} + \hat{d}_{i}^{\dag}({\bf
p},\zeta)v({\bf p},\zeta)e^{ip \cdot x}] \,\, ,
\end{equation}
where
\begin{equation}
d\tilde{p}_{i} = \frac{d^{3}{\bf p}}{(2\pi)^{3}2E_{i}(p)}
\end{equation}
 is a Lorentz-invariant measure and $p \cdot x = p_\mu x^\mu = Et-{\bf p}\cdot{\bf x}$.

Eq.~(\ref{dirac_field}) is a solution of the free Dirac equation
$[i\gamma^{\mu}\partial_{\mu} - m_{i}]\hat{\psi}_{i}(x) = 0$. The
operator $\hat{b}^{\dag}_{i}({\bf p},\zeta)$ creates a neutrino of
mass $m_{i}$, momentum {\bf p} and spin polarization $\zeta$, while
$\hat{d}^{\dag}_{i}({\bf p},\zeta)$ creates an antineutrino of mass
$m_{i}$, momentum {\bf p} and spin polarization $\zeta$. The
operators $\hat{b}_{i}({\bf p},\zeta)$ and $\hat{d}_{i}({\bf
p},\zeta)$ annihilate the respective particles. The particle states
belong to the Hilbert space $\mathcal{H}$, which can be taken to be
the Fock space constructed on the vacuum state $|0\rangle$ defined
by the condition $\hat{b}_{i}({\bf p},\zeta)|0\rangle =
\hat{d}_{i}({\bf p},\zeta)|0\rangle = 0$. The spinors $u({\bf
p},\zeta)$ and $v({\bf
p},\zeta)$ are defined by
\begin{equation}\label{u}
u({\bf p},\zeta) = \sqrt{E_{i}(p)+m_i}\left( \begin{array}{c}
 \chi({\bf p},\zeta)\\
 \frac{{\bf \sigma}\cdot{\bf p}}{E_i(p)+m_i}\chi({\bf p},\zeta)\\
 \end{array} \right),
\end{equation}
\begin{equation}\label{v}
v({\bf p},\zeta) = \sqrt{E_{i}(p)+m_i}\left( \begin{array}{c}
 \frac{{\bf \sigma}\cdot{\bf p}}{E_i(p)+m_i}\chi({\bf p},-\zeta)\\
 \chi({\bf p},-\zeta)
 \end{array} \right).
\end{equation}
Here $\chi({\bf p},\zeta)$ is a two-component spinor of momentum ${\bf p}$ and fixed polarization $\zeta$ which we take to be independent of $m_i$ and ${\bf p}$. The spinor $\chi({\bf p},\zeta)$ is normalized according to
\begin{equation}\label{weyl_normalization}
\chi^{\dag}({\bf p},\zeta)\chi({\bf p},\eta) = \delta_{\zeta\eta}.
\end{equation}
From Eqs.~(\ref{u}~-~\ref{weyl_normalization}) we get
\begin{equation} \label{spinor_normalization}
\bar{u}({\bf p},\zeta)\gamma^{\mu}u({\bf p},\eta) = 2p^{\mu}\delta_{\zeta\eta},
\end{equation}
\begin{equation}
\bar{v}({\bf p},\zeta)\gamma^{\mu}v({\bf p},\eta) = 2p^{\mu}\delta_{\zeta\eta},
\end{equation}
\begin{equation}
\bar{u}({\bf p},\zeta)\gamma^{0}v({\bf -p},\eta) = 0.
\end{equation}

The fundamental anticommutation relations between the fields $\hat{\psi}_{i}({\bf x},t)$
and their canonical momenta $\hat{\Pi}_{i}({\bf x},t) =
i\hat{\psi}^{\dag}_{i}({\bf x},t)$ are:
\begin{equation}
\{\hat{\psi}_{i}({\bf x},t),\hat{\Pi}_{j}({\bf y},t)\} =
\{\hat{\psi}_{i}({\bf x},t),i\hat{\psi}^{\dag}_{j}({\bf y},t)\} =
i\delta^{3}({\bf x} - {\bf y})\delta_{ij}\,\, ,
\end{equation}
\begin{equation}
\{\hat{\psi}_{i}({\bf x},t),\hat{\psi}_{j}({\bf y},t)\} = 0 \,\, ,
\,\,\, \{\hat{\Pi}_{i}({\bf x},t),\hat{\Pi}_{j}({\bf y},t)\} =
0\,\,.
\end{equation}
{}From these equations we derive the anticommutation relations between
the operators of creation $\hat{b}^{\dag}_{i}({\bf p},\zeta)$ and
annihilation $\hat{b}_{i}({\bf p},\zeta)$
\begin{equation}\label{anticommutation}
\{ \hat{b}_{i}({\bf p},\zeta),\hat{b}_{j}^{\dag}({\bf q},\eta) \} =
(2\pi)^{3}2E_{i}(p)\delta^{3}({\bf p} - {\bf
q})\delta_{\zeta\eta}\delta_{ij}\,\,,
\end{equation}
\begin{equation}
\{\hat{b}_{i}({\bf p},\zeta), \hat{b}_{j}({\bf q},\eta) \} = 0 \,\,,
 \,\,\, \{\hat{b}^{\dag}_{i}({\bf p},\zeta),
\hat{b}^{\dag}_{j}({\bf q},\eta) \} = 0\,\,.
\end{equation}
Similar relations with $\hat{b}$ replaced by $\hat{d}$ hold for the
antiparticle creation and annihilation operators
$\hat{d}_{i}^{\dag}({\bf p},\zeta)$ and $\hat{d}_{i}({\bf
p},\zeta)$.

A Majorana field is described by Eq. (\ref{dirac_field}) with the
additional relations
\begin{equation} \label{Majorana_field}
\psi_{i}(x) = C\overline{\psi_{i}}^{T}(x)\,\,,\,\,\,\hat{d}_{i}({\bf
p},\zeta)=\hat{b}_{i}({\bf p},\zeta)\,\,\, ,
\end{equation}
where $C =
i\gamma^{0}\gamma^{2}$ is the charge conjugation matrix. To facilitate the transition to a Majorana field, we choose the
phases of $u({\bf p},\zeta)$ and $v({\bf p},\zeta)$ so that $v({\bf
p},\zeta)$ equals the charge conjugate of $u({\bf p},\zeta)$,
namely $v({\bf p},\zeta) = C \bar{u}^{T}({\bf p},\zeta)$.
The quantized Majorana field is thus explicitly
\begin{equation}
\hat{\psi}_{i}(x) = \sum_{\zeta}\int d\tilde{p}_{i}[\hat{b}_{i}({\bf
p},\zeta)u({\bf p},\zeta)e^{-ip \cdot x} + \hat{b}^{\dag}_{i}({\bf
p},\zeta)C \bar{u}^{T}({\bf p},\zeta)e^{ip \cdot x} ]\,\,.
\end{equation}
The normalization of the $\hat{b}_{i}({\bf p},\zeta)$ and
$\hat{b}^{\dag}_{i}({\bf p},\zeta)$ operators is the same as in Eq.~(\ref{anticommutation}).

We now discuss the states of the system. In the Heisenberg picture free states
do not evolve in time \cite{srednicki}. A free state $|{\bf p},\zeta,i\rangle$ of momentum ${\bf p}$,
polarization $\zeta$ and mass $m_{i}$ is given by
\begin{equation} \label{state}
|{\bf p},\zeta,i\rangle = \frac{\hat{b}_{i}^{\dag}({\bf
p},\zeta)}{\sqrt{(2\pi)^{3}2E_{i}(p)}}|0\rangle\,\,.
\end{equation}
The factor in the denominator comes from our choice of normalization
condition
\begin{equation} \label{normalizzation}
\langle{\bf p},\zeta,i|{\bf q},\eta,j\rangle = \delta^{3}({\bf p} -
{\bf q})\delta_{\zeta\eta}\delta_{ij} \,\,,
\end{equation}
which we use for both Dirac and Majorana particles.

The states $|{\bf p},\zeta,i\rangle$ satisfy the eigenvalue
equations
\begin{eqnarray}
&& \hat{P}^{\mu}|{\bf p},\zeta,i\rangle = p^{\mu}|{\bf
p},\zeta,i\rangle\,\,,
\\ &&
\hat{N}|{\bf p},\zeta,i\rangle = |{\bf p},\zeta,i\rangle\,\,,
\\ &&
\hat{S}_n|{\bf p},\zeta,i\rangle = \zeta |{\bf p},\zeta,i\rangle\,\,,
\end{eqnarray}
where the four-momentum operators $\hat{P}^{\mu}$, the number operator
$\hat{N}$, and the spin operator projected onto the polarization axis $\hat{S}_n$ are respectively
\begin{eqnarray} &&
\hat{P}^{\mu} = \sum_\zeta \int d\tilde{p}\, p^{\mu}\,[\hat{b}^{\dag}_{i}({\bf
p},\zeta)\hat{b}_{i}({\bf p},\zeta) + \hat{d}^{\dag}_{i}({\bf
p},\zeta)\hat{d}_{i}({\bf p},\zeta)]\,\,,
\\ &&
\hat{N} = \sum_\zeta \int d\tilde{p}\,[\hat{b}^{\dag}_{i}({\bf
p},\zeta)\hat{b}_{i}({\bf p},\zeta) + \hat{d}^{\dag}_{i}({\bf
p},\zeta)\hat{d}_{i}({\bf p},\zeta)]\,,
\\ &&
\hat{S}_n = \sum_\zeta \int d\tilde{p}\, \zeta \,[\hat{b}^{\dag}_{i}({\bf
p},\zeta)\hat{b}_{i}({\bf p},\zeta) + \hat{d}^{\dag}_{i}({\bf
p},\zeta)\hat{d}_{i}({\bf p},\zeta)]\,,
\end{eqnarray}
if the states are described by the Dirac theory, or
\begin{eqnarray} &&
\hat{P}^{\mu} = \sum_\zeta \int d\tilde{p}\, p^{\mu}\,\hat{b}^{\dag}_{i}({\bf
p},\zeta)\hat{b}_{i}({\bf p},\zeta)\,\,,
\\ &&
\hat{N} = \sum_\zeta \int
d\tilde{p} \,\hat{b}^{\dag}_{i}({\bf p},\zeta)\hat{b}_{i}({\bf
p},\zeta)\,,
\\ &&
\hat{S}_n = \sum_\zeta \int d\tilde{p}\, \zeta\,\hat{b}^{\dag}_{i}({\bf
p},\zeta)\hat{b}_{i}({\bf p},\zeta)\,,
\end{eqnarray}
in the Majorana theory.

As a consequence of Eq.~(\ref{normalizzation}), the states $|{\bf
p},\zeta,i\rangle$ belong to a continuum spectrum and are not
normalizable. Following standard procedure, we introduce wave
packets in Fock space. Let
\begin{equation} \label{wave_packet}
|c_{i}\rangle = \sum_{\zeta}\int d^{3}{\bf p}\,a({\bf
p},\zeta)\,|{\bf p},\zeta,i\rangle .
\end{equation}
This is a wave packet specified by the function $a({\bf p},\zeta)$, which represents the
probability amplitude in momentum space for the state
$|c_{i}\rangle$. Notice that wave packets in the Heisenberg representation do not depend on time. The wave packet is normalized according to
\begin{equation} \label{norm}
\langle c_{i}|c_{i}\rangle = \sum_{\zeta}\int d^{3}{\bf p} |a({\bf
p},\zeta)|^{2} = 1\,\,.
\end{equation}
Another way to see that $a({\bf p},\zeta)$ is the momentum
probability amplitude is the expression of the momentum expectation
value
\begin{equation}
 \langle c |\hat{P}^{\mu}|c\rangle = \sum_\zeta \int d^{3} {\bf p}\, p^{\mu} \,|a({\bf
 p},\zeta)|^{2}\,\,.
\end{equation}
This expression applies for both Dirac and Majorana particles.

We will discuss the use of the momentum-space distributions $a({\bf p},\zeta)$ to derive the probability amplitude in Section~\ref{Oscillation formula for Dirac neutrinos}.

The first-quantized wave function corresponding to the free
wave packet $ |c_{i}\rangle $ in Eq.~(\ref{wave_packet}) can be
computed from Eq.~(\ref{dirac_field}) and (\ref{state}) to be
\begin{equation} \label{wave_function}
\psi_{i}({\bf x},t) = \langle 0|\hat{\nu}_{i}(x)|c_{i}\rangle =
\sum_\zeta \int \frac{d^{3}{\bf p}}{\sqrt{(2\pi)^{3}
2E_{i}(p)}}e^{i{\bf p}\cdot{\bf x}-iE_i({\bf p})t} a({\bf p},\zeta)u_i({\bf
p},\zeta)\,\,.
\end{equation}
The wave function $\psi_i({\bf x},t)$ satisfies the Dirac equation for any choice of the momentum-space probability amplitude $a({\bf p},\zeta)$. The relativistic wave function $\psi_i({\bf x},t)$ is not the probability amplitude in position space. The latter, in the interpretation of Newton and Wigner \cite{NW,FG}, is given by the Fourier transform of $a({\bf p},\zeta)$,
\begin{equation}\label{a_tilde}
\tilde{a}({\bf x},\zeta) = \int \frac{ d^3{\bf p} }{ (2\pi)^{3/2} } \, a({\bf p},\zeta) e^{i{\bf p}\cdot{\bf x}}.
\end{equation}

Finally we recall the expression of the inner product between two
one-particle states in Fock space as expressed in terms of wave functions. Let
the one-particle wave packets $|c_{1}\rangle$ and $|c_{2}\rangle$,
corresponding to momentum-space amplitudes $a_1({\bf p},\zeta)$ and
$a_2({\bf p},\zeta)$, be represented by the wave functions
$\psi_{1}(x)$ and $\psi_{2}(x)$ as in Eq.~(\ref{wave_function}).
Then the equal-time inner product can be written in several equivalent forms:
\begin{eqnarray} \label{dotproduct}
\langle c_{2} | c_{1} \rangle & = & \sum_\zeta \int d^3{\bf p} \,
a_{2}^*({\bf p},\zeta) \, a_1({\bf p},\zeta) = \sum_\zeta \int d^3{\bf x} \,
{\tilde a}_{2}^*({\bf x},\zeta) \, {\tilde a}_1({\bf x},\zeta) \nonumber \\ & = & \int d^3{\bf x} \,
\overline{\psi}_{2}({\bf x},t) \gamma^0 \psi_{1}({\bf x},t) = \int d^3{\bf x} \,
{\psi}_{2}^\dagger({\bf x},t) \psi_{1}({\bf x},t).
\end{eqnarray}
Notice that for a non-interacting
field the value of this inner product is independent of time.

\section{Theory of Dirac and Majorana neutrinos}

Given a four-component spinor $\psi(x)$ it is always possible to project out
its left-handed (LH) and right-handed (RH) parts $\psi_{L}(x)$ and
$\psi_{R}(x)$ by means of the projection operators $P_{L} = \frac{1
- \gamma^{5}}{2}$ and $P_{R} = \frac{1 + \gamma^{5}}{2}$, respectively. Thus,
\begin{equation}
\psi_{L}(x) = P_{L}\psi(x) \,\,,\,\, \psi_{R}(x) =
P_{R}\psi(x)\,\,.
\end{equation}

In the following we indicate a LH field for a neutrino of
flavor $\alpha$ as $\nu_{L,\alpha}(x)$. It is this field that enters
the Lagrangian of the Standard Model.

For Dirac neutrino fields, a mass term cannot be
described by using the $\nu_{L,\alpha}(x)$ field component only but a RH
field $\nu_{R,\alpha}(x)$ is needed as well. The term describing Dirac neutrino masses is~\cite{BP}
\begin{equation} \label{mass_term}
\mathcal{L}^{D}_{m} = \sum_{\alpha\beta}
\overline{\nu}_{R,\alpha}M_{\alpha\beta}\nu_{L,\beta} + {\rm h.c.}\,\,,
\end{equation}
where $M_{\alpha\beta}$ is the mass matrix.

In order to diagonalize $M_{\alpha\beta}$, one introduces two unitary
matrices $U$ and $V$ \cite{BP, KM}, in terms of which
\begin{equation}
M_{\alpha\beta} = \sum_{i} V_{\alpha
i}m_{i}U^{*}_{\beta i}\,\,.
\end{equation}
The quantity $m_{i}$ is the mass of the $i$-th eigenstate. The flavor eigenstates are rotated into the neutrino mass
eigenstates $\nu_{L,i}$ and $\nu_{R,i}$ as
\begin{equation} \label{rotation}
\nu_{L,i} = \sum_{\alpha} U_{\alpha i}^{*}\nu_{L,\alpha} \,\,,\,\,\, \nu_{R,i} = \sum_{\alpha}
V_{\alpha i}^{*}\nu_{R,\alpha}\,\,.
\end{equation}
The mass term becomes
\begin{equation}
\mathcal{L}_{m} = \sum_{i} m_{i}\overline{\nu}_{R,i}\nu_{L,i} + {\rm h.c.} = \sum_{i}
m_{i}\overline{\nu}_{i}\nu_{i} \,\,.
\end{equation}
Here $\nu_{i} = \nu_{R,i}+\nu_{L,i}$.
For future application we use the following parametrization for the two flavor generation mixing matrix $U$
\begin{equation}
U = \left( \begin{array}{cc}
 \cos\theta & \sin\theta\\
 -\sin\theta & \cos\theta\\
 \end{array} \right).
\label{parametrization}
\end{equation}

For Majorana neutrinos, it is possible to construct a mass term in the Lagrangian by
using just one chirality, say LH, for the field $\nu$. The most general Majorana mass
term is
\begin{equation} \label{mass_term1}
\mathcal{L}^{M}_{m} =
\frac{1}{2}\,\sum_{\alpha,\beta}\,(\nu_{L,\alpha})^{C}M_{\alpha\beta}\nu_{L,\beta}
+ {\rm h.c.}\,\,.
\end{equation}
Here $(\nu_{L})^{C} = C \overline{\nu_{L}}^{T}$ is the charge-conjugate of $\nu_{L}$, and it is a RH field.
In this case, in order to diagonalize the mass term in Eq.~(\ref{mass_term1}), only one unitary matrix $W_{\alpha i}$ is needed,
\begin{equation}
M_{\alpha\beta} = \sum_{i}W_{i \alpha}^{*}m_{i}W_{\beta i}^{*}.
\end{equation}
The rotated basis
describing the left-handed mass eigenstates is
\begin{equation}
\nu_{L,i} =
\sum_{\alpha}W^{*}_{i \alpha}\nu_{L,\alpha}.
\end{equation}
In the new basis, the Lagrangian term in Eq.~(\ref{mass_term1}) becomes
\begin{equation}
\mathcal{L}^{M}_{m} = \frac{1}{2}\sum_{i}m_{i}
(\nu_{L,i})^{C}\nu_{L,i} + {\rm h.c.} =
\frac{1}{2}\sum_{i}m_{i}\overline{\chi}_{i}\chi_{i} .
\end{equation}
Here $\chi_{i} = \nu_{L,i} + (\nu_{L,i})^{C}$ is a Majorana field.

The mixing matrix $W$ of $N$ Majorana neutrinos contains $N-1$
physical phases in addition to the angles in the mixing matrix $U$ of
the Dirac theory (see e.g.~\cite{BP}). This is due to the fact
that the Majorana mass term in Eq.(\ref{mass_term1}) is not
invariant under the U(1) global symmetry $\nu_{L,\alpha} \to
e^{-i\phi_{\alpha}}\nu_{L,\alpha}$.

Following \cite{eidelman} it is possible to write the
Majorana mixing matrix $W$ as the product $W=UD$ of a unitary matrix $U$ similar in form to
the Dirac mixing matrix in Eq.~(\ref{parametrization}) and a diagonal unitary matrix $D =
\mathop{\rm diag}(1,e^{i\phi_2},e^{i\phi_3},\ldots)$:
\begin{equation}
W_{\alpha i} = U_{\alpha i} e^{i \phi_i} .
\label{parametrization2}
\end{equation}
For the two-flavor case, one has explicitly
\begin{equation}
W = \left( \begin{array}{cc}
 \cos\theta & \sin\theta e^{i\phi}\\
 -\sin\theta & \cos\theta e^{i\phi}\\
 \end{array} \right).
\label{parametrization1}
\end{equation}

\begin{figure}[t]
  \centering
  \includegraphics[width=4cm]{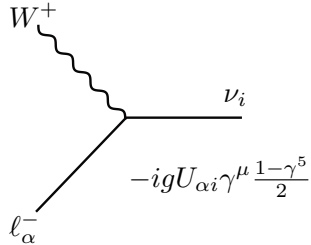}
  \caption{The Feynman vertex describing the weak interaction between a charged lepton $l_{\alpha}$, a charged vector boson $W_{\mu}$ and a neutrino mass eigenstate $\nu_{i}$ of mass $m_{i}$. The formula next to the diagram gives the strength of the coupling and depends on the mixing matrix $U_{\alpha i}$.}\label{Figure_1}
\end{figure}

Finally, we recall what happens to the weak interaction vertices
because of the change of basis from interaction to mass eigenstates. We take as an example the lagrangian for the weak interaction of neutrinos,
charged leptons and $W$ gauge bosons
\begin{equation}
\mathcal{L}_{int} =
g\overline{\nu}_{L,\alpha}\gamma^{\mu}l_{L,\alpha}W_{\mu} + {\rm h.c.}.
\end{equation}
In the last expression, $\nu_{L,\alpha}$ is a LH neutrino field, which can be Dirac or Majorana, $l_{L,\alpha}$ is the LH spinor field for the charged lepton of flavor $\alpha$, $W_{\mu}$ is the charged gauge boson field and $g$ is the weak coupling strength. In the mass eigenstate basis, this
lagrangian becomes
\begin{equation} \label{vertex}
\mathcal{L}_{int} = g\overline{\nu}_{L,i}U_{\alpha
i}\gamma^{\mu}l_{L,\alpha}W_{\mu} + {\rm h.c.}\,\,.
\end{equation}
The
vertex coupling a lepton $l_{L,\alpha}$ of flavor $\alpha$ to a
neutrino $\nu_i$ of a given mass can be read off Eq. (\ref{vertex})
as $-igU_{\alpha i}\gamma^{\mu}\frac{1-\gamma^{5}}{2}$. This vertex
is shown in Fig.~\ref{Figure_1}.

\section{Oscillation formula for Dirac neutrinos} \label{Oscillation formula for Dirac neutrinos}

We now consider the following experiment, which resembles the actual operation of long-baseline neutrino experiments using accelerator beams. Imagine a burst of neutrinos created with a specific flavor $\alpha$ by some mechanism of production. We are not interested in the characteristics of the production process because, following standard quantum mechanics, once the initial conditions are given the propagation of the wave packet is determined. The source is fixed in the laboratory frame of coordinate ${\bf X}_P$ and the burst occurs at time $t_P$ in this frame. We set a detector at the point ${\bf X}_{D}$ and we gather information about the neutrino flavor $\beta$ at time $t_D=t_P+\Delta t$. If desired, we could turn on the detector for a certain time $\Delta T_{det}$, in which case w would integrate the oscillation probability over the time interval $\Delta T_{det}$. Again we do not take into account any theory concerning the mechanism of detection, but only assume we want the probability that the neutrino will be in a given final state at time $t_D$. Because of the phenomenon of neutrino oscillations, the flavor $\beta$ of the neutrino burst at detection can differ from the flavor at production $\alpha$. We are interested in computing the probability to observe the flavor $\beta$ at detection if the neutrino had a specific flavor $\alpha$ at production. Our final formula depends on the time delay $\Delta t$ and on the characteristics of the mass eigenstate wave functions in momentum space $a_i({\bf p})$; it will be valid in general for any shape of the wave functions.

In Section \ref{gaussian_wave_packets} we specialize the wave functions $a_i({\bf p})$ assuming Gaussian distributions with average momenta ${\bf P}_i$ and dispersions $\sigma_{p}$. The dispersion in momentum space is related to the uncertainty in the localization of the source (or detection point) by the relation $\sigma_{p} \sigma_{x} = 1/2$. 

We now proceed to derive the neutrino oscillation probability $\mathcal{P}_{\beta\alpha}$ using a complete quantum field theory treatment. Following Eq~(\ref{wave_packet}) we take the wave function describing the mass eigenstate of mass $m_i$ at production as
\begin{equation}\label{wave}
\nu_{i,P}(x) = \sum_{\zeta_P}\int \frac{d^{3}{\bf p}}{\sqrt{(2\pi)^{3}2E_{i}(p)}}a_{i,P}({\bf
p},\zeta_{P})u({\bf p},\zeta_{P})e^{-ip\cdot x}\,\,.
\end{equation}
Here the extra label ($P$) indicates the production state.
The wave function describing the flavor $\alpha$ is the linear combination of the wave functions Eq.~(\ref{wave})
\begin{equation}
\nu_{\alpha,P}(x) = \sum_i U_{\alpha i}\nu_{i,P}(x).
\end{equation}
We use the Dirac propagator $S_{i}(x - y)$ to evolve the wave function Eq.~(\ref{wave}) from the production time $t_P$ to the detection time $t_D = t_P + \Delta t$. The wave function Eq.~(\ref{wave}) evolves as
\begin{equation} \label{evoluted1}
\nu_{i,P}(x) = i\int d^{3}{\bf y}S_{i}(x-y)\gamma^{0}\nu_{i,P}(y).
\end{equation}
The propagator $S_{i}(x - y)$ is given by \cite{srednicki}
\begin{equation} \label{propagator1}
S_{i}(x- y) = \int \frac{d^{4}k}{(2\pi)^{4}}\frac{\gamma^{\mu}k_{\mu}+m_{i}}{k^{\mu}k_{\mu}-m^{2}_{i}+i\epsilon}e^{-ik^{\mu}(x-y)_{\mu}}.
\end{equation}
As in the laboratory frame the time delay is $\Delta t$, we rewrite Eq.~(\ref{propagator1}) as
\begin{equation} \label{propagator}
S_{i}(\Delta t,{\bf x}- {\bf y}) = \int \frac{d^{4}k}{(2\pi)^{4}}\frac{\gamma^{\mu}k_{\mu}+m_{i}}{k^{\mu}k_{\mu}-m^{2}_{i}+i\epsilon}e^{-ik_{0}\Delta t + i{\bf k}\cdot({\bf x} - {\bf y})}.
\end{equation}
We integrate over $k_0$ using
\begin{equation}
\int dk_0 \frac{e^{-ik_{0}\Delta t}}{(k_{0}-E_{k}+i\epsilon)(k_{0}+E_{k}-i\epsilon)} = -2\pi i\frac{e^{-iE_{i}({\bf k})\Delta t}}{2E_{i}({\bf k})}\,\,.
\end{equation}
where we assumed that $\Delta t>0$ so the integration contour encloses the region $\Im k_{0} > 0$. This accounts for the propagation of particles to be forward in time. This gives
\begin{equation}
S_{i}(\Delta T,{\bf x}- {\bf y}) = -i\int \frac{d^{3}{\bf k}}{(2\pi)^{3}}\frac{\gamma^{\mu}k_{\mu}+m_{i}}{2E_i(k)}e^{-iE_{i}(k)\Delta t + i{\bf k}\cdot({\bf x} - {\bf y})}.
\end{equation}
Notice that $k_0\Delta t$ in Eq.~(\ref{propagator}) has now become $E_i(k)\Delta t$. The evolved wave function follows as
\begin{equation}
\nu_{i,P}(t+\Delta t,{\bf x}) = i\int d^{3}{\bf y}S_{i}(\Delta t,{\bf x} - {\bf y})\gamma^{0}\nu_{i,P}(t,{\bf y}) =\nonumber
\end{equation}
$$=\sum_{\zeta_P}\int d^{3}{\bf k}\,d^{3}{\bf p}\,a_{i,P}({\bf
p},\zeta_{P})\frac{(\gamma^{\mu}k_{\mu}+m_i)\gamma^{0}u({\bf p},\zeta_{P})}{(2\pi)^3 2E_i(k)\sqrt{(2\pi)^{3}2E_i(p)}}\,e^{-iE_{i}(k)\Delta t -iE_i(p)t+i{\bf k}\cdot{\bf x}}\int d^{3}{\bf y}\,e^{-i{\bf y}\cdot({\bf k}-{\bf p})} = $$
\begin{equation} \label{evoluted2}
= \sum_{\zeta_P}\int d^{3}{\bf p}\,a_{i,P}({\bf
p},\zeta_{P})\frac{(\gamma^{\mu}p_{\mu}+m_i)\gamma^{0}u({\bf p},\zeta_{P})}{2E_i(p)\sqrt{(2\pi)^{3}2E_i(p)}}\,e^{-iE_{i}(p)(t+ \Delta t)+i{\bf p}\cdot{\bf x}}.
\end{equation}
In the last step we used $\int d^{3}{\bf y}\,e^{-i{\bf y}\cdot({\bf k}-{\bf p})} = (2\pi)^3 \delta({\bf k}-{\bf p})$. Now the following spinor relations hold
\begin{equation}
\gamma^{\mu}p_{\mu}\gamma^{0}u({\bf p},\zeta_P) = \left[-\gamma^{0}m +2E_{i}(p)\right] u({\bf p},\zeta_P),
\end{equation}
and
\begin{equation}
(\gamma^{\mu}p_{\mu} + m_i)\gamma^0 u({\bf p},\zeta_P) = 2E_i(p)u({\bf p},\zeta_P).
\end{equation}
This gives the neutrino wave function after a time $\Delta t$ as
\begin{equation}\label{evoluted}
\nu_{i,P}(t+\Delta t,{\bf x}) = \sum_{\zeta_P} \int \frac{d^{3}{\bf
p}}{\sqrt{(2\pi)^{3}2E_{i}(p)}}a_{i,P}({\bf p},\zeta_P)u({\bf
p},\zeta_P)e^{-iE_i(p)(t +\Delta t)+i{\bf p}\cdot{\bf x}}\,\,,
\end{equation}
which is obtained from Eq.~(\ref{wave}) by letting $t \to t+\Delta t$.

Notice from this detailed calculation that the propagator rightly contributes only the time dependent factor $e^{-iE_i(p)\Delta t}$, which is the matrix element of the evolution operator in the energy representation. There is no extra term $e^{i {\bf p} \cdot ({\bf X}_D - {\bf X}_P)}$. The latter term will appear in the production and detection amplitudes when the respective wave packets are centered around the production and detection points.

We now define the wave function centered around the detection point for mass eigenstate $m_i$:
\begin{equation} \label{wave2}
\nu_{i,D}(x) = \sum_{\zeta_D} \int \frac{d^{3}{\bf p}}{\sqrt{(2\pi)^{3}2E_{i}(p)}}a_{i,D}({\bf
p},\zeta_{D})u({\bf p},\zeta_{D})e^{-iE_i(p)t+i{\bf p}\cdot{\bf x}}.
\end{equation}
The wave function describing the neutrino wave packet with flavor $\beta$ at detection is
\begin{equation}
\nu_{\beta,D}(x) = \sum_i U_{\beta i}\nu_{i,D}(x).
\end{equation}

The probability amplitude for the neutrino oscillations is defined as
\begin{equation} \label{amplitude1}
\mathcal{A}_{\alpha\beta} = \int d^{3}{\bf x}
\bar{\nu}_{\beta,D}(t,{\bf x})\gamma^{0}\nu_{\alpha,P}(t+\Delta
t,{\bf x}) = \sum_{i,j}U^{*}_{\beta j}U_{\alpha i}\int d^{3} \, {\bf
x}\bar{\nu}_{j,D}(t,{\bf x})\gamma^{0}\nu_{i,P}(t+\Delta t,{\bf x}).
\end{equation}
Inserting Eqs~(\ref{evoluted}) and~(\ref{wave2}) in Eq.~(\ref{amplitude1}), the spinor normalization in Eq.~(\ref{spinor_normalization}) gives $\zeta_P =\zeta_D\equiv\zeta$, and we obtain
\begin{equation} \label{amplitude2a}
\mathcal{A}_{\beta\alpha} =
\sum_{i,\zeta}U^{*}_{\beta i}U_{\alpha i}\int
d^{3}{\bf p}\, a_{i,D}^{*}({\bf p},\zeta)a_{i,P}({\bf
p},\zeta)e^{-iE_{i}(p)\Delta t}.
\end{equation}

Eq.~(\ref{amplitude2a}) is our general result, obtained using basic principles of quantum mechanics only. No approximation was used to derive this equation, which is valid in general for any form of the wave packets $a_{i,\rho}({\bf p},\zeta)$. The production and detection mechanisms fix the form of the amplitudes $a_{i,P}({\bf p},\zeta)$ and $a_{i,D}({\bf p},\zeta)$, respectively. Eq.~(\ref{amplitude2a}) gives the right probability amplitude with the correct normalization when $\Delta t = 0$ and $P = D$. In this case the integrand reduces to $|a_{i,P}({\bf p},\zeta)|^2d^3{\bf p}$ which is the probability of finding a particle in the infinitesimal volume of the momentum space $d^3{\bf p}$ centered at ${\bf p}$.

We now discuss the properties of the neutrino oscillation probability $\mathcal{P}_{\beta\alpha}$. This is given by
\begin{equation}
\mathcal{P}_{\beta\alpha} = |\mathcal{A}_{\beta\alpha}|^2 = \left|\sum_{i,\zeta} U^{*}_{\beta i}U_{\alpha i}\int
d^{3}{\bf p}\, a_{i,D}^{*}({\bf p},\zeta)a_{i,P}({\bf
p},\zeta)e^{-iE_{i}(p)\Delta t}\right|^{2} .
\label{eq:Pmasseigenstates}
\end{equation}
An interesting relation is obtained for the sum $\sum_\beta \mathcal{P}_{\beta\alpha}$, which represents the probability of observing a neutrino after time $\Delta t$ independently of its flavor. Using Eq.~(\ref{eq:Pmasseigenstates}) and the unitarity of the mixing matrix $U$, one finds
\begin{equation}
\sum_\beta \mathcal{P}_{\beta\alpha} = \sum_{i,\zeta} |U_{\alpha i}|^2 \mathcal{P}_i,
\label{eq:sumofP}
\end{equation}
where $\mathcal{P}_i$ is the probability of observing the $i$-th mass eigenstate if there were no mixing between neutrinos,
\begin{equation}
\mathcal{P}_i = \left|\sum_{\zeta} \int
d^{3}{\bf p}\, a_{i,D}^{*}({\bf p},\zeta)a_{i,P}({\bf
p},\zeta)e^{-iE_{i}(p)\Delta t}\right|^{2}
\end{equation}
Eq.~(\ref{eq:sumofP}) states that the probability $\sum_\beta \mathcal{P}_{\beta\alpha}$ of observing a neutrino in any flavor after time $\Delta t$ equals the weighted average of the probabilities $\mathcal{P}_i$ of observing a neutrino in each mass eigenstate, weighted by the probability $|U_{\alpha i}|^2$ of being in that eigenstate in the initial state.

Eq.~(\ref{eq:sumofP}) applies in general, for coherent and non-coherent cases, and for oscillations or lack of oscillations. As illustrated in Section~\ref{sec:v} below, it allows a clear understanding of the phenomenon of decoherence, and of the question of the arrival time of the neutrino.

\section{Oscillation formula for Majorana neutrinos}

Here we compute the probability amplitude for Majorana neutrinos and review the well-known result that there is no difference from the Dirac theory \cite{BP,giunti1}.
The flavor-eigenstate wave packets are
\begin{equation}
|c_{\alpha,\rho}\rangle = \sum_{i}W_{\alpha i}|c_{i,\rho}\rangle =
\sum_{i,\zeta_{\rho}} W_{\alpha i}\int d^{3}{\bf p}\,a_{\rho}({\bf
p},\zeta_{\rho})\,|{\bf p},\zeta_{\rho},i\rangle \,\,,
\end{equation}
where $\rho = P,D$ labels the production and detection points and the states $|{\bf p},\zeta,i\rangle$ are now defined in terms
of the creation operators $\hat{b}^{\dag}_{i}({\bf p},\zeta)$ of the
Majorana theory. The flavor wave functions are
\begin{equation}\label{wave1}
\chi_{\alpha,\rho}(x) = \sum_{i}W_{\alpha i}\langle
0|\hat{\nu}_{i}(x)|c_{i,\rho}\rangle = \sum_{i,\zeta_{\rho}}W_{\alpha i}\int
\frac{d^{3}{\bf p}}{\sqrt{(2\pi)^{3}2E_{i}(p)}}\,a_{1,\rho}({\bf p},\zeta_{\rho})u({\bf
p},\zeta_{\rho})e^{-ip\cdot x}\,\,.
\end{equation}
Following the same steps in the derivation of the oscillation
formula Eq.~(\ref{amplitude2a}) for Dirac neutrinos, we arrive to the following analogous expression in
the Majorana case:
\begin{eqnarray}\label{amplitude4}
\lefteqn{ \mathcal{A}_{\alpha\beta} = \int d^{3}{\bf x}
\chi^{\dag}_{\beta,2}(x)\chi_{\alpha,1}(x) = \sum_{i}W^{*}_{\beta
i}W_{\alpha i}\int d^{3} \, {\bf
x}\chi^{\dag}_{i,2}(x)\chi_{i,1}(x)=} \\
&& =
\sum_{i,\zeta}W^{*}_{\beta i}W_{\alpha i}\int
d^{3}{\bf p} a_{i,D}^{*}({\bf p},\zeta)a_{i,P}({\bf
p},\zeta)e^{-iE_{i}(p)\Delta t}.
\end{eqnarray}
When the matrix $W$ is parametrized as in Eq.~(\ref{parametrization2}), one finds
\begin{equation}
W^{*}_{\beta i}W_{\alpha i} = (U_{\beta i} e^{i\phi_i})^* (U_{\alpha i} e^{i\phi_i}) = U^{*}_{\beta i}U_{\alpha i}.
\end{equation}
Thus
\begin{equation}
\mathcal{A}_{\alpha\beta} = \sum_{i,\zeta}U^{*}_{\beta i}U_{\alpha i}\int
d^{3}{\bf p}\, a_{i,D}^{*}({\bf p},\zeta)a_{i,P}({\bf
p},\zeta)e^{-iE_{i}(p)\Delta t}.
\end{equation}
This expression is identical to Eq.~(\ref{amplitude2a}) for Dirac neutrinos.
The physical phases $\phi_i$ play no role in the theory of neutrino oscillations and it is impossible to distinguish between Dirac and Majorana neutrinos on the basis of their oscillations phenomena (see e.g. \cite{BP,giunti1}).

\section{Gaussian wave packets} \label{gaussian_wave_packets}

A minimal-dispersion Gaussian wave packet with average position ${\bf X}$, average momentum ${\bf P}$, spin polarization $\zeta$ and spatial dispersion $\sigma_x$ is represented at time $t=0$ by
\begin{equation}
{\tilde a}_{{\bf X},{\bf P},\xi,\sigma_x}({\bf x},\zeta) = \frac{\delta_{\zeta \xi}}{(2\pi
\sigma_{x}^{2})^{3/4}}e^{-\frac{({\bf x}-{\bf
X})^{2}}{4\sigma_{x}^{2}}}e^{i{\bf P}\cdot{\bf x}}
\,\,,
\end{equation}
where $\delta_{\zeta \xi}$ is the Kronecker symbol which fixes the polarization along an axis $\xi$. Its momentum-space probability amplitude is a Gaussian with dispersion $\sigma_{p} =
1/(2\sigma_{x})$:
\begin{equation}
a_{{\bf X},{\bf P},\zeta,\sigma_p}({\bf p},\zeta) = \frac{\delta_{\zeta\xi}}{(2\pi
\sigma_{p}^{2})^{3/4}}e^{-\frac{({\bf p}-{\bf
P})^{2}}{4\sigma_{p}^{2}}}e^{-i{\bf p}\cdot{\bf X}} \,\, .
\end{equation}
Notice that the term $e^{-i{\bf p}\cdot{\bf X}}$ is a function of momentum, not position, and arises from displacing a wave packet whose average position is at the origin into one centered at ${\bf X}$.

We now specialize the production amplitude by assuming that the distribution in momentum space $a_{i,P}({\bf p},\zeta_P)$ is represented by a minimal-dispersion Gaussian wave packet with average momentum ${\bf P}_{i,P}$, average position ${\bf X}_P$, spin polarization along a fixed direction $\zeta_P$ and momentum dispersion $\sigma_{p,P}$:
\begin{equation} \label{distribution}
a_{i,P}({\bf p},\zeta_P) = a_{{\bf X},{\bf P},\zeta,\sigma_p}({\bf p},\zeta).
\end{equation}
A similar equation holds for $a_{i,D}({\bf p},\zeta_D)$ by replacing $P \to D$. The Fourier transforms of these packets are centered at the production (${\bf X}_P$) and the detection (${\bf X}_D$) point, respectively.

As discussed in \cite{giunti,giunti1} Gaussian wave packets are just approximations of the true distributions, whose shapes will depend on the details of the production and detection processes.

Using the distribution~(\ref{distribution}) in momentum space we get the following probability amplitude for the process
\begin{equation} \label{amplitude2}
\mathcal{A}_{\beta\alpha} =
\delta_{\zeta_{D}\zeta_{P}}\sum_{i}U^{*}_{\beta i}U_{\alpha i}\int
\frac{d^{3}{\bf p}}{(2\pi\sigma_{pD}\sigma_{pP})^{3/2}}
e^{-\frac{({\bf p}-{\bf P}_{i,D})^{2}}{4\sigma^{2}_{pD}}-\frac{({\bf
p}-{\bf P}_{i,P})^{2}}{4\sigma^{2}_{pP}}-iE_{i}(p)\Delta t + i{\bf
p}\cdot\Delta {\bf X}}.
\end{equation}
The dependence on the production and detection points ${\bf X}_P$ and ${\bf X}_D$ is now explicit because of the choice of the Gaussian distribution~(\ref{distribution}). In Eq.~(\ref{amplitude2}) we introduced the notation $\Delta {\bf X} = {\bf X}_{D} -
{\bf X}_{P}$.

Combining the quadratic terms in the exponential, one obtains the known result (see e.g. \cite{giunti}) that the relevant momentum dispersion $\sigma^2_{pPD}$ is a combination of the momentum dispersions at production and detection,
\begin{equation}
\frac{1}{\sigma^2_{pPD}} = \frac{1}{\sigma^2_{pD}} +\frac{1}{\sigma^2_{pP}} .
\end{equation}
The separation of the wave packets will be governed by the combined dispersion $\sigma_{pPD}$. In particular, the spatial coherence of the source, as parametrized by $\sigma_{pP}$, enters the determination of wave packet separation.

In the illustrative case in which the initial and final
polarizations, the average momenta, and the momentum dispersions are
the same, i.e.\ $\zeta_{D} = \zeta_{P}$, ${\bf P}_{i,D} = {\bf P}_{i,P}
\equiv {\bf P}_i$ and $\sigma_{pD} = \sigma_{pP} \equiv \sigma_p$,
we have
\begin{equation} \label{amplitude3}
\mathcal{A}_{\beta\alpha} = \sum_{i}U^{*}_{\beta i}U_{\alpha i}\int
\frac{d^{3}{\bf p}}{(2\pi\sigma_p^{2})^{3/2}}e^{-\frac{({\bf p}-{\bf
P}_i)^{2}}{2\sigma_p^2} - iE_{i}(p)\Delta t + i{\bf
p}\cdot\Delta{\bf X}} .
\end{equation}
This is the basic expression we use below to illustrate neutrino
oscillations and decoherence.

We now expand the energy term $E_i(p)$ appearing in Eq.(\ref{amplitude3}) to second order in the momentum ${\bf q}= {\bf p}
- {\bf P}_i$. We find
\begin{equation} \label{expansion}
E_i(p) = E_i(P_i) +
\frac{\vec{p}\cdot\vec{q}}{E_i(P_i)} +
\frac{m_i^2}{E_i(P_i)^3}|\vec{q}|^2 + O(|\vec{q}|^3).
\end{equation}
This expansion is a good approximation when $\frac{{\bf P}_i\cdot{\bf
q}}{E_i(P_i)} \ll E_i(P_i)$. Since the momentum distribution is
Gaussian with dispersion $\sigma_p$, we have $|\vec{q}| \lesssim
\sigma_p$ and the expansion in Eq.~(\ref{expansion}) applies for
\begin{equation}
\sigma_p \ll P_i + \frac{m_i^2}{P_i},
\end{equation}
where $P_i = |{\bf P}_i|$. The condition is valid for
non-relativistic neutrinos (for which $\sigma_p\ll m_i$ but $\sigma_p$ can be smaller or larger than $P_i$), while in the ultra-relativistic limit it
is equivalent to $\sigma_p \ll P_i$. This conclusion differs from that in \cite{giunti1}, where this approximation is said to be valid only in
the ultra-relativistic limit.

The first order term in Eq.~(\ref{expansion}) is responsible for
separation of the wave packets, while the second term is related to their spreading. To see this, we integrate Eq.~(\ref{amplitude3}) with the expansion~(\ref{expansion}) up to second order and find
\begin{equation} \label{second_order}
\mathcal{A}_{\beta\alpha} = \sum_{i}U^{*}_{\beta i}U_{\alpha
i}\frac{\sigma^3_p(t)}{\sigma^3_p}e^{-\frac{\sigma_p^2(t)}{2}({\bf v}_i\Delta
t - \Delta{\bf X})^2-iE_{i}(P_i)\Delta t + i{\bf P_i}\cdot\Delta{\bf
X}}.
\end{equation}
Here
\begin{equation}
{\bf v}_i = \frac{{\bf P}_i}{E_i(P_i)}
\end{equation}
is the group
velocity of the $i$-th mass eigenstate, and
\begin{equation}
\sigma^2_p(t) =
\frac{\sigma^2_p}{1+\omega^2(t)}\left[1-i\omega(t)\right] ,
\end{equation}
where
\begin{equation}
\omega(t) =
\frac{\sigma_p^2 m_i^2\Delta t}{(m_i^2+P_i^2)^{3/2}}.
\end{equation}
The quantity $\sigma_p(t)$ is the
time-dependent dispersion of the wave packet in momentum space, in analogy with the spreading of a Schrodinger-like wave packet
\cite{messiah}.

At small $\Delta t \ll {(m_i^2+P_i^2)^{3/2}}/({\sigma_p^2 m_i^2})$ (for
all $i$), the spreading of the wave packets can be neglected, $\sigma_p(t) \approx \sigma_p$. Then the oscillation amplitude and the detection
probability become
\begin{equation} \label{amplitude_ultrarel}
\mathcal{A}_{\beta\alpha} = \sum_{i}U^{*}_{\beta i}U_{\alpha
i}e^{-\frac{1}{8\sigma_x^2}({\bf v}_i\Delta
t - \Delta{\bf X})^2-iE_{i}(P_i)\Delta t + i{\bf P_i}\cdot\Delta{\bf
X}},
\end{equation}
\begin{equation} \label{probability_ultrarel}
\mathcal{P}_{\beta\alpha} = \sum_{ij}U^*_{\beta i}U_{\alpha
i}U_{\beta j}U^*_{\alpha j}e^{i({\bf P}_i-{\bf P}_j)\cdot\Delta {\bf
X}-i[E_i(P_i)-E_j(P_j)]\Delta t} e^{-\frac{1}{8\sigma_x^2}[({\bf
v}_i\Delta t - \Delta{\bf X})^2+({\bf v}_j\Delta t - \Delta{\bf
X})^2]}.
\end{equation}
Here $\sigma_x = 1/2\sigma_p$ is the dispersion in position space.
In these formulas the product of the $U$'s and the oscillating
exponential are the usual expressions for plane waves, and the last
exponential is the envelope describing the superposition of the two
Gaussian wave packets.

In Eqs.~(\ref{second_order}) and~(\ref{amplitude_ultrarel}), each mass eigenstate $i$ contributes a phase
\begin{equation}
\Phi_i = E_i(P_i) \Delta t - {\bf P}_i \cdot \Delta {\bf X} .
\end{equation}
We want to point out that the quantity $\Delta t$ appearing here is \emph{not} the time of flight of the neutrino mass eigenstate $i$. Rather it is the time delay between production and observation of the neutrino packet. This time delay is fixed by the experimental situation, and is independent
of the neutrino mass eigenstate. On the contrary, the time of flight for the $i$-th mass eigenstate is given by $\Delta t_i = \Delta X/v_i$, and is different for different masses. If one were incorrectly to replace $\Delta t$ with $\Delta t_i$, one would obtain a spurious factor of 2 in the oscillation formula (see e.g.~\cite{giunti1}, Section 8.4.2, and references therein). Our expression does not contain such a spurious factor of two. Moreover, it is relativistically invariant (if the spreading of the wave packet is neglected).

\section{Two neutrino flavors}\label{two_neutrino_flavors}

In the following we illustrate the probability in Eq.~(\ref{probability_ultrarel}) by considering two neutrino flavors only, e.g.\ $\mu$ and $\tau$ flavors, neglecting the mixing with the electronic flavor or with other hypothetical sterile neutrinos. Imagine a muon neutrino produced by some mechanism at a source located at ${\bf X} = {\bf X}_P$. This muon neutrino is described as a superposition of two neutrino mass eigenstates wave packets of mass $m_1$ and $m_2$ with average momenta ${\bf P_1}$ and ${\bf P_2}$, respectively, starting at the same time $t = 0$ with the same average position ${\bf X}$. In this scenario we can compute from Eq.~(\ref{probability_ultrarel}) the probability $\mathcal{P}_{\mu\to\tau}$ of observing a tau neutrino after a time $\Delta t$ with a detector placed at distance $\Delta {\bf X}$ from the production point. Using the parametrization in Eq.~(\ref{parametrization}), the oscillation probability  formula in Eq.~(\ref{probability_ultrarel}) specializes to
\begin{equation} \label{probability_2gen}
\mathcal{P}_{\mu\to\tau} = \sin^2(2\theta) \, \left[ \left( \frac{z_1-z_2}{2} \right)^2 +z_1 z_2 \sin^2\left(\frac{\Delta E \Delta t - \Delta {\bf P}\cdot\Delta {\bf X}}{2}\right)\right].
\end{equation}
As explained before, the same expression is obtained for both Dirac and Majorana neutrinos.

In Eq.~(\ref{probability_2gen}) we have defined
\begin{equation}
z_i \equiv e^{-\frac{1}{8\sigma_x^2}(\Delta {\bf X}- {\bf v}_i\Delta
t)^2}
\end{equation}
Physically, $z_i^2$ is the overlap integral between the initial and the final Gaussian wave packets for the $i$-th mass eigenstate when the spreading of the wave packet is neglected, as in Eq.~(\ref{probability_ultrarel}). Notice that $z_i = 1$ at $\Delta {\bf x}=\Delta t=0$ and on the classical trajectory ${\Delta {\bf X}} = {\bf v}_i {\Delta t}$, and $z_i<1$ elsewhere. The product $z_1 z_2 \sin^2(2\theta)$ is the overlap integral between states of the same flavor at production and detection. Oscillations are present only when the latter overlap integral is different from zero.

The probability to observe the same flavor $\mu$ can be similarly obtained as
\begin{equation} \label{probability_2gen1}
\mathcal{P}_{\mu\to\mu} = \left( z_1^2 \cos^2\theta + z_2^2 \sin^2\theta \right)^2 - z_1 z_2 \sin^2(2\theta) \sin^2\left(\frac{\Delta E \Delta t - \Delta {\bf P}\cdot\Delta {\bf X}}{2}\right).
\end{equation}
From the last two equations we see that the total probability of
detection of any flavor, if we started with the flavor $\mu$, is
\begin{equation}
\mathcal{P}_{\mu} \equiv
\mathcal{P}_{\mu\to\tau}+\mathcal{P}_{\mu\to\mu} = z_1^2\cos^2\theta + z_2^2\sin^2\theta.
\label{total_prob}
\end{equation}
This probability is one when $\Delta {\bf x}=\Delta t=0$ and when computed along the classical trajectory
${\Delta {\bf X}} = {\bf v}_i {\Delta t}$. Away from the classical trajectory, the probability of observing a neutrino is
smaller than one. This may seem to be an unusual result, since one would expect that the probability of observing one of the two flavors has to be 100\%. However, what we are computing here is not the probability that if we start with a neutrino at the origin we observe it to be somewhere at a later time, which correctly is one. Instead, we are computing the probability that an initial Gaussian neutrino wave function centered at the origin at time $t=0$ has become a final Gaussian of the same dispersion centered at $\Delta {\bf X}$ at a later time $\Delta t$. This probability is given by the overlap integrals $z_i^2$ between the initial and the final Gaussians.  Eq.~(\ref{total_prob}) for the total probability of observing any flavor at a later time is then easily interpreted as the sum of the probability of producing and detecting each mass eigenstate, eigenstate $1$ being produced with probability $\cos^2\theta$ and observed with probability $z_1^2$, eigenstate $2$ being produced with probability $\sin^2\theta$ and observed with probability $z_2^2$. This is a special case of the general result expressed by Eq.~(\ref{eq:sumofP}).

In the plane wave limit, $\sigma_p\to 0$, $\sigma_x\to \infty$, and $z_i\to 1$. Eqs.~(\ref{probability_2gen}--\ref{total_prob}) reduce to
\begin{eqnarray}
\label{eq:60}
\mathcal{P}_{\mu\to\tau} &= &\sin^2(2\theta) \, \sin^2\!\left(\frac{\Delta E \Delta t - \Delta {\bf P}\cdot\Delta {\bf X}}{2}\right),
\\
\mathcal{P}_{\mu\to\mu} & = & 1 - \sin^2(2\theta) \, \sin^2\!\left(\frac{\Delta E \Delta t - \Delta {\bf P}\cdot\Delta {\bf X}}{2}\right),
\\
\mathcal{P}_{\mu} &=&1.
\label{eq:62}
\end{eqnarray}
Eqs.~(\ref{eq:60}--\ref{eq:62}) contain the relativistically invariant combination $\Delta E \Delta t - \Delta {\bf P}\cdot\Delta {\bf X} = \Delta P_\mu \Delta X^\mu$, where four-vector notation has been introduced.

In the ultra-relativistic limit $P_i \approx E - m_i^2/(2E)$ where $E \approx E_1 \approx E_2$ to the first order. In this limit, $\Delta E=0$, $\Delta P = \Delta m^2/(2E)$, and
\begin{equation}
\mathcal{P}_{\mu\to\tau} = \sin^2(2\theta) \sin^2\left( \frac{\Delta m^2 \Delta X}{4E} \right)
\label{eq:sigma0gen}
\end{equation}
\begin{equation}
\mathcal{P}_{\mu\to\mu} = 1 - \sin^2(2\theta) \sin^2\left( \frac{\Delta m^2 \Delta X}{4E} \right).
\label{eq:sigma0gen1}
\end{equation}
In an alternative derivation of the ultra-relativistic limit, we could set $\Delta t \approx \Delta X$ and use $\Delta E \approx \Delta P + \Delta m^2/(2E)$, which follows from $P_i \approx E_i + m^2_i/(2E_i)$. We obtain $\Delta P_\mu \Delta X^\mu/2 \approx \Delta E (\Delta t - \Delta X)/2 + \Delta m^2 \Delta X/(4E) \approx \Delta m^2 \Delta X/(4E) $. Again this has the correct factor of 4 in the denominator.

However we remark that Eqs. (\ref{eq:60}--\ref{eq:62}) hold without assuming $\Delta E=0$. In particular, one does not need to assume that the energy or the momentum of the different wave packets are equal. Indeed, such assumptions are not Lorentz invariant, and would be valid in a particular Lorentz frame only. Depending on the particular process in which neutrinos are produced, there will exist a frame in which it is possible to equate either the bulk momenta or the energies of the eigenstates, but this makes little importance in our treatment as we already chose our reference frame by setting the relative distance of source and detector $\Delta {\bf X}$ and the time delay $\Delta t$.

It is convenient to define a coherence length between two wave packets. Their average position at time $t_P+\Delta t$ is given by ${\bf X}_i = {\bf X}_P + {\bf v}_i \Delta t$, where ${\bf
v}_i = {\bf P}_i/\sqrt{m_{i}^{2}+P_i^2}$ is the wave packet group
velocity. We define the coherence length $L_{\rm coh}$ as the
distance $|({\bf X}_1+{\bf X}_2)/2-{\bf X}_P|$ from the source at
which the two wave packets are separated by an amount equal to twice
their spatial dispersion $2\sigma_{x}$, that is $|{\bf v}_1-{\bf
v}_2| \Delta t = 2 \sigma_x$. The coherence length follows easily as
\begin{equation} \label{coherence_length}
L_{\rm coh} = \frac{|{\bf v}_{1}+{\bf v}_2|}{|{\bf v}_{1}-{\bf
v}_{2}|}\, \sigma_x.
\end{equation}
The coherence length sets the distance from the source within which it is possible to observe oscillations, since it is gives the value of the overlap integral $z_1 z_2 \sin^2(2\theta)$ (see also Figure~(\ref{separating}) below). As pointed out in \cite{stodolsky} the coherence length is proportional to $\sigma_x$, the spatial dispersion at production and at detection. It is thus related to the characteristics of productions and the resolution of the detector. As we stress in this paper, we are not interested in these details as once they are set the properties of the oscillation probability are determined.

\section{Numerical illustration - ultra-relativistic neutrinos}
\label{sec:v}

In this section we show the time behavior of the oscillation
probability $\mathcal{P}_{\beta\alpha}$ by means of a numerical illustration, limiting the treatment to the case of two flavor generations.
We consider atmospheric neutrinos created by the interaction of cosmic rays with the Earth atmosphere. We take the following values for the parameters in Eqs.~(\ref{probability_2gen}--\ref{probability_2gen1}),
\begin{equation}
\Delta X = 6000\,{\rm km}, \quad \Delta m^{2} = 2.7 \times
10^{-3}\,{\rm eV}^{2}/c^4, \quad E = 2\,{\rm GeV} , \quad \theta =
45^\circ , \quad \sigma_{x} = 10\,{\rm km}. \label{eq:table1}
\end{equation}
We imposed the same energy $E$ for both neutrinos. Since $E \gg m_i$, the ultra-relativistic limit applies. In this limit one has $v_i \simeq 1 - m_i^2/(2E^2)$, and the coherence length $L_{\rm coh}$ becomes
\begin{equation}
L_{\rm coh} \simeq \frac{4E^2}{\Delta m^2}\sigma_x\,\,(\hbox{for}\,\,E \gg m_{1,2}).
\end{equation}
For the values in Eqs.~(\ref{eq:table1}), one finds $L_{\rm coh} \simeq 6 \times 10^{22} \, {\rm km} \simeq 2 \, {\rm Gpc} $, comparable to the Hubble radius. So, oscillations are coherent on the scale of the Earth.

For $\Delta X \ll L_{\rm coh}$, one can set $z_2 = z_1 \equiv z$, and the overlap integral $z^2$ factorizes in the oscillation formulas,
\begin{equation}\label{our_amplitude1}
\mathcal{P}_{\mu\to\tau} = z^2\sin^2(2\theta) \sin^2\left(\frac{\Delta m^2\Delta X}{4E}\right),
\end{equation}
\begin{equation}\label{our_amplitude2}
\mathcal{P}_{\mu\to\mu} = z^2 \left[1-\sin^2(2\theta) \sin^2\left(\frac{\Delta m^2\Delta X}{4E}\right)\right].
\end{equation}

\begin{figure}[t]
  \centering
  \includegraphics[height=7cm,width=10cm]{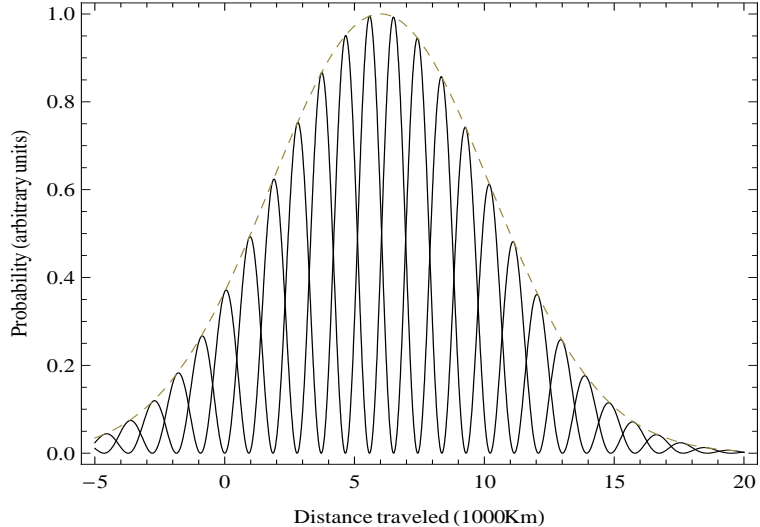}
  \caption{An example of the probability of detecting a neutrino in a Gaussian state when produced in a Gaussian state, plotted as a function of the distance from the source $\Delta X$ at a specific time $\Delta t$ after production. We assume two neutrino generations with masses and mixing angles given in Eq.~\ref{eq:table1} and take $\Delta t = 20$ ms. The dashed line shows the probability of detecting either flavor: this probability peaks at $\Delta X = c\Delta t = 6000\, {\rm km}$ corresponding to the classical trajectory for ultra-relativistic neutrinos. The two intertwined solid lines show the probability of detecting one or the other neutrino flavor; oscillations are clearly visible. }\label{Fig_amplitude}
\end{figure}

An important feature of the oscillation phenomenon in the
ultra-relativistic limit is its dependence on $\Delta m^2$ only. This fact makes an oscillation experiment insensitive to the individual values of the masses $m_1$ and $m_2$.

{}From Eqs.~(\ref{our_amplitude1}--\ref{our_amplitude2}), the oscillation length is
\begin{equation}\label{l_osc}
L_{\rm osc} = 4\pi\frac{E}{\Delta m^2}.
\end{equation}
(This is the distance between two consecutive peaks in the probability of observing a certain flavor.)
In this relativistic limit, the ratio of the coherence and oscillation lengths is independent of $\Delta m^2$:
\begin{equation}
\frac{L_{\rm coh}}{L_{\rm osc}} \simeq
\frac{E}{8\pi\sigma_p}\,\,\hbox{for}\,\,E \gg m_{1,2}\,\,.
\end{equation}
This ratio is just the relative momentum dispersion $\sigma_p/E$ that determines the number of oscillation cycles before coherence is lost and the two particles wave packets separate in space.

In Fig.~\ref{Fig_amplitude}, we plot $\mathcal{P}_{\mu\to\tau}$ and $\mathcal{P}_{\mu\to\mu}$ in Eqs.~(\ref{our_amplitude1}--\ref{our_amplitude2}) as functions of the distance $\Delta X$ at time $\Delta t =  20$ ms, for the parameters given in Eq.~\ref{eq:table1}. The two intertwined solid lines show the probabilities of observing one or the other neutrino flavor. The maxima are spaced by a length equal to $L_{\rm osc} = 1.86\times10^{3}$ km, as per Eq.~(\ref{l_osc}). The dashed line shows the probability $\mathcal{P}_{\mu\to\tau}+\mathcal{P}_{\mu\to\mu}$ of detecting either flavor. This total probability is the overlap integral $z^2$ and it peaks on the classical trajectory $\Delta X = c \Delta t = $ 6000 km. As the neutrinos are ultra-relativistic, the probabilities of detecting different mass eigenstates overlap and it is not possible to distinguish between the two packets in this limit.

\section{Numerical illustration - decoherence}

We now discuss the phenomenon of neutrino decoherence in the case where the distance between the source and the detector is $\Delta X \gtrsim L_{\rm coh}$. For this purpose we focus again on a two-generation model, with neutrino flavors $\mu$ and $\tau$, but we choose an admittedly non-realistic example of non-relativistic neutrinos created at the source with the same average momenta $P_1 = P_2 = P$.  The parameters are as follows
\begin{eqnarray}
& m_1 = 2\cdot 10^{-3} {\rm eV}/c^2,
\quad m_2 = 0.022 {\rm eV}/c^2, \quad P =
2.2\cdot 10^{-4}{\rm eV}/c , & \nonumber \\
& \Delta X = 6000 {\rm km}, \quad \theta=45^\circ, \quad
\sigma_{x} = 10 {\rm km}. & \label{eq:table2}
\end{eqnarray}
The oscillation probability is described by Eqs.~(\ref{probability_2gen}--\ref{probability_2gen1}).

In the non-relativistic limit, the coherence length from Eq.~(\ref{coherence_length}) becomes
\begin{equation}
L_{\rm coh} = \frac{2(m_1+m_2)^2}{\Delta m^2}\sigma_x,
\end{equation}
while the oscillation length is
\begin{equation}
L_{\rm osc} = 2\pi \frac{m_1+m_2}{\Delta m^2}.
\end{equation}

\begin{figure}[t]
  \centering
  \includegraphics[height=7cm,width=10cm]{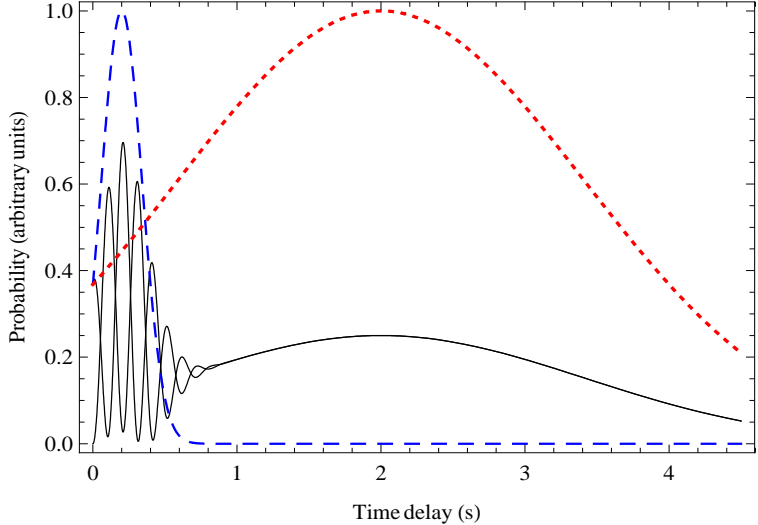}
  \caption{The neutrino oscillation probability
  Eqs.~(\ref{probability_2gen}), (\ref{probability_2gen1})
as a function of the time delay $\Delta t$ between neutrino
production and detection, in the case of two-neutrino generations
and for a coherence length $L_{coh}$ comparable to the distance
$\Delta X$ between the production and detection points. The
parameters used for the plot are listed in Eq.~(\ref{eq:table2}).
The dashed (blue) and dotted (red) lines represent the probabilities to observe
the mass eigenstates of lighter or heavier masses $m_1$ or
$m_{2}$,respectively. Notice that the lighter neutrino arrives before the heavier one. The solid black lines show the probability
that the initial flavor $\mu$ is detected as $\tau$ or $\mu$ after a
time $\Delta t$. Oscillations occur only when the lighter and heavier neutrino wave packets overlap.}\label{Fig_amplitude1}
\end{figure}

We have chosen the parameters in
Eq.~(\ref{eq:table2}) so that $L_{\rm coh} \simeq \Delta X$.

As can be seen in Fig.~\ref{Fig_amplitude1}, oscillations are suppressed at late times, when the two wave packets do no longer overlap. The solid lines represent $\mathcal{P}_{\mu\to\tau}$ and $\mathcal{P}_{\mu\to\mu}$, the transition probabilities of observing neutrinos of flavor $\tau$ or $\mu$, respectively, if the initial flavor is $\mu$. These probabilities are bounded by
\begin{eqnarray} \label{probability_bound}
& \frac{\sin^2(2\theta)}{4} (z_1-z_2)^2 < \mathcal{P}_{\mu\to\tau} < \frac{\sin^2(2\theta)}{4}(z_1+z_2)^2 & \\
& \frac{\sin^2(2\theta)}{4} (z_1-z_2)^2 < \mathcal{P}_{\mu\to\mu} < \frac{\sin^2(2\theta)}{4}(z_1+z_2)^2. &
\end{eqnarray}
As $v_1 > v_2$, the probabilities of detecting a neutrino of flavor $\mu$ or $\tau$ drops as $\frac{1}{4}\sin^2(2\theta) z_1^2$ for $\Delta t \gtrsim {\Delta X}/({v_1 - v_2})$.

The separation of the wave packets is clear in Fig.~\ref{Fig_amplitude1}. If there were no mixing, since the two packets have the same bulk momentum $P$, the lighter wave packet (dashed blue bell-like curve on the left, peaking at $\sim0.2$ s) would arrive sooner than the heavier one (dotted red bell-like curve on the right, peaking at $\sim2$ s). With mixing, the heights of the peaks are suppressed in the manner of Eq.~(\ref{eq:sumofP}): the light-neutrino peak is suppressed by a factor $\cos^2\theta$ and the heavy-neutrino peak by a factor $\sin^2\theta$. Oscillations can occur only when the packets have a substantial overlap (out to $\sim0.8$ s in our example). A detector with timing capabilities placed at $\Delta X \gtrsim L_{\rm coh}$ would be able to distinguish the two mass eigenstates based on their different times of flight from the production point. The ability to discern the two mass eigenstates destroys the interference pattern, in a way similar to the famous double slit experiment.

\begin{figure}[h!]
  \centering
  \includegraphics[height=7cm,width=17cm]{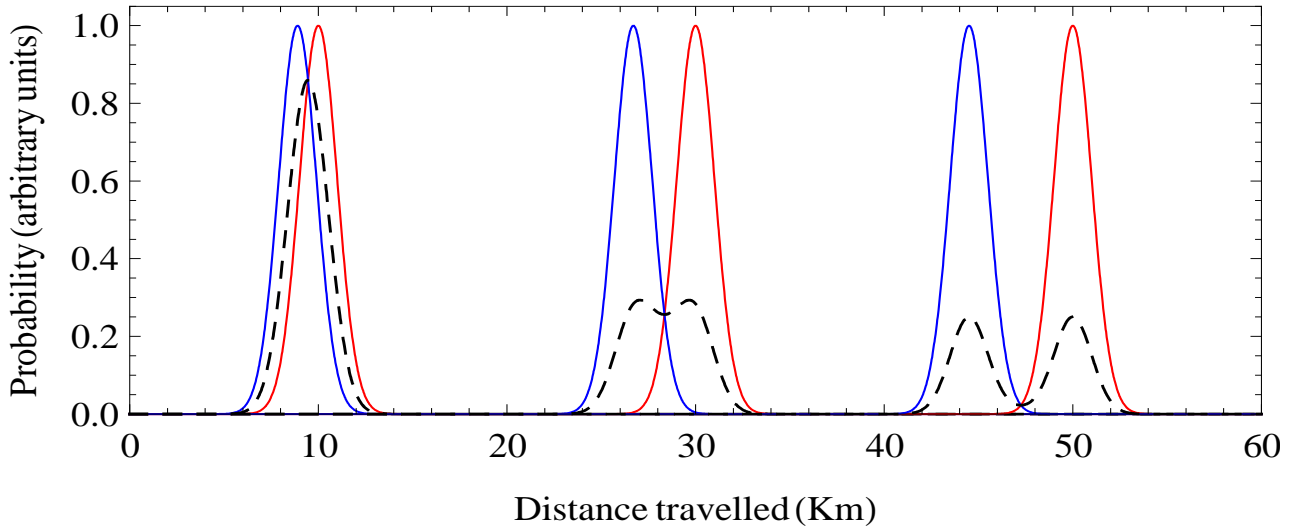}
  \caption{Illustration of how two wave packets produced at the same point of space-time with different group velocities gradually diminish their overlap losing coherence. Three snapshot at successive times are shown (from left to right): $\Delta t$ = 10s, 25s and 50s. The solid lines show the probability of observing one or the other of the mass eigenstates. Notice how the peaks of the probabilities separate. The dashed lines show the envelopes of the flavor oscillation probabilities (not shown because they vary on a very short distance and would blur the figure if shown in their entirety).}
  \label{separating}
\end{figure}

Another way of seeing the separation of the wave packets is to plot the detection probability as a function of the distance from the source at different times from production. We do this in Figure~\ref{separating} for three different times. To produce a clear illustration, we have chosen the following parameters:
\begin{eqnarray}
& m_1 = 0.1 {\rm eV}/c^2,
\quad m_2 = 0.11 {\rm eV}/c^2, \quad P =
10^{-3}{\rm eV}/c ,
 \quad \theta=45^\circ, \quad
\sigma_{x} = 10 {\rm km}, &
\end{eqnarray}
and $\Delta t=$ 10 s, 25 s, and 50 s. The first snapshot at $\Delta t=10$ s shows the two wave packets (solid blue and red lines) overlapping. As a consequence the probability of detecting a neutrino of any flavor is relatively high, as shown by the dashed line representing the envelope of the oscillation probabilities (not shown). The second snapshot at $\Delta t=25$ s shows the wave packets when they are separated by a spatial distance equal to $\sigma_x$, which is half of the coherence length as defined by us. In the third snapshot, at $\Delta t=50$ s, the wave packets are no longer overlapping; the probabilities of detecting a neutrino flavor mirror those of the mass eigenstates, but are suppressed by the probability of producing the specific flavor in the first place.

\section{Discussion and conclusions}

In this paper we have derived the probability of detecting a
neutrino with a specific flavor $\beta$ by using basic rules of
standard quantum mechanics. In doing this we only used properties
belonging to the neutrino mass eigenstates $\nu_i$, the flavor
eigenstates being just a linear combination of $\nu_i$ according to
the mixing matrix $U$ (if Dirac) or $W$ (if Majorana). We
gave physical reasons about the choice of using the mass basis instead of the flavor basis, as the former diagonalizes the mass
matrix in the hamiltonian. For the probability amplitude in position
space we used the Fourier transform of the momentum probability
$a({\bf p})$, as in the prescription by Newton and Wigner. This allowed us to find the correct
expression for the oscillation amplitude, Eq.(\ref{amplitude3}).

We commented on the equal energy prescription often found in the literature on neutrino oscillations. Thanks to Eqs.~(\ref{probability_2gen}) and (\ref{probability_2gen1}), which express the oscillation probability in a Lorentz invariant way, we showed that this requirement is not necessary and oscillations are generally attained also when the energies and the momenta of the mass eigenstates are not equal. In the two-generation case, it may be possible to choose a Lorentz frame in which the neutrino energies are equal. In the general case of more than two flavors, this may not always be possible, depending on the details of neutrino production. If the production process is stationary, equal energies can be assumed (see \cite{stodolsky}), but in general it is not assured that a specific frame can be found where all of the energies are equal.

We have also stressed that the oscillation amplitude
depends on the details of the production and detection
processes only through the initial conditions for the flavor
eigenstate wave packets. To illustrate the physics of the neutrino
oscillations, we have shown some numerical examples for a
two-dimensional flavor space. In particular, we have presented a
simple explanation of the separation of the wave packets and the consequent loss of coherence of the oscillations.

\begin{acknowledgments}
The authors would like to thank C.~Giunti and L.~Stodolsky for reading the manuscript and for helpful suggestions and discussions. This work was supported by NSF grants PHY-0456825 and PHY-58501336 at the University of Utah.
\end{acknowledgments}

\end{document}